\documentclass[aps,prd,twocolumn,superscriptaddress,nofootinbib]{revtex4-1}

\usepackage{latexsym}
\usepackage{amsmath}
\usepackage{amssymb}
\usepackage{amsfonts}
\usepackage{bm}
\RequirePackage{lineno}

\usepackage{color}
\definecolor{purple}{rgb}{0.5,0,0.5}
\definecolor{blue}{rgb}{0.0,0,0.9}
\definecolor{prdblue}{rgb}{0.133,0.118,0.498}
\usepackage[colorlinks=true, pdfstartview=FitV, linkcolor=prdblue, citecolor= prdblue, urlcolor=prdblue]{hyperref}

\usepackage{supertabular} 
\usepackage{placeins}
\usepackage{epsfig}
\usepackage{graphicx}
\usepackage{hyperref}

\hypersetup{
  breaklinks=true,
  colorlinks = true,
  linkcolor = blue,
  anchorcolor = blue,
  citecolor = blue,
  filecolor = blue,
  pagecolor = blue,
  urlcolor = blue
}
\hypersetup{
  bookmarks=true,
  bookmarksnumbered=true,
  bookmarkstype=toc,
  linktocpage=true
}

\begin{document}

\modulolinenumbers[2]

\setlength{\oddsidemargin}{-0.5cm} \addtolength{\topmargin}{15mm}

\title{\boldmath Study of $\Lambda_c^+\rightarrow \Lambda \mu^+\nu_{\mu}$ and Test of Lepton Flavor Universality with $\Lambda_c^+\rightarrow \Lambda \ell^+\nu_{\ell}$ Decays}
\author{
  \small
M.~Ablikim$^{1}$, M.~N.~Achasov$^{13,b}$, P.~Adlarson$^{75}$, X.~C.~Ai$^{81}$, R.~Aliberti$^{36}$, A.~Amoroso$^{74A,74C}$, M.~R.~An$^{40}$, Q.~An$^{71,58}$, Y.~Bai$^{57}$, O.~Bakina$^{37}$
% ...
   \\
   \vspace{0.2cm}
   (BESIII Collaboration)\\
   \vspace{0.2cm} {\it
$^{1}$ Institute of High Energy Physics, Beijing 100049, People's Republic of China\\
% ...
$^{13}$ G.I. Budker Institute of Nuclear Physics SB RAS (BINP), Novosibirsk 630090, Russia\\
% ...
$^{36}$ Johannes Gutenberg University of Mainz, Johann-Joachim-Becher-Weg 45, D-55099 Mainz, Germany\\
$^{37}$ Joint Institute for Nuclear Research, 141980 Dubna, Moscow region, Russia\\
% ...
$^{40}$ Liaoning Normal University, Dalian 116029, People's Republic of China\\
% ...
$^{57}$ Southeast University, Nanjing 211100, People's Republic of China\\
$^{58}$ State Key Laboratory of Particle Detection and Electronics, Beijing 100049, Hefei 230026, People's Republic of China\\
% ...
$^{71}$ University of Science and Technology of China, Hefei 230026, People's Republic of China\\
% ...
$^{74}$ University of Turin and INFN, (A)University of Turin, I-10125, Turin, Italy; (B)University of Eastern Piedmont, I-15121, Alessandria, Italy; (C)INFN, I-10125, Turin, Italy\\
$^{75}$ Uppsala University, Box 516, SE-75120 Uppsala, Sweden\\
% ...
$^{81}$ Zhengzhou University, Zhengzhou 450001, People's Republic of China\\
\vspace{0.2cm}
% ...
$^{b}$ Also at the Novosibirsk State University, Novosibirsk, 630090, Russia\\
% ...
   \vspace{0.4cm}
}
}

\begin{abstract}
The measurement of the Cabibbo-favored semileptonic decay
$\Lambda_c^+\rightarrow \Lambda \mu^+\nu_{\mu}$ is reported using $4.5~\mathrm{fb}^{-1}$ of $e^+e^-$
annihilation data collected at center-of-mass energies ranging from 4.600~GeV to 4.699~GeV.
The branching fraction of the decay is measured to be 
$\mathcal{B}(\Lambda_c^+\rightarrow \Lambda \mu^+\nu_{\mu})=(3.48\pm0.14_{\rm stat.}\pm0.10_{\rm syst.})\%$, three times more precise than the prior world average result. 
Tests of lepton flavor universality using $\Lambda_c^+\rightarrow \Lambda \ell^+\nu_{\ell}$ ($\ell=e, \mu$) decays are reported for the first time, based on measurements of the differential decay rates and the forward-backward asymmetries in separate four-momentum transfer regions. The results are compatible with Standard Model predictions.  
Furthermore, we improve the determination of the form-factor parameters in $\Lambda_c^+\rightarrow \Lambda \ell^+\nu_{\ell}$ decays, which provide stringent tests and calibration for lattice quantum chromodynamics (LQCD) calculations.

\end{abstract}

\maketitle
Semileptonic (SL) decays play a critical role in the determination of the Cabibbo-Kobayashi-Maskawa (CKM) matrix~\cite{CKM} elements which are the fundamental parameters of the Standard Model (SM). The SM predicts that the electroweak interactions have the identical strengths for all the three different lepton generations, which is known as lepton flavor universality (LFU).
In recent years, tests of LFU show some hints of tension with SM predictions in both tree-level $b\rightarrow c\ell^{\prime}\nu_{\ell^{\prime}}$ ($\ell^{\prime}=\tau, e, \mu$) transitions~\cite{pdg2020,PRL109_101802,PRD88_072012,PRD92_072014,PRD94_072007,PRL115_159901,PRL118_21101,PRL120_171802,PRD97_072013,PRL120_171802} and flavor-changing neutral current $b\rightarrow s\ell^+\ell^-$ ($\ell=e, \mu$) transitions~\cite{PRD86_032012,JHEP06_108,JHEP08_055,PRL122_191801,JHEP05_040,JHEP03_105,NP18_277}. 
Thus far, the experimental results on the combination of $\mathcal{R}(D)$ and $\mathcal{R}(D^*)$ show a discrepancy with respect to the SM prediction of 
about $3.4\sigma$~\cite{EPJC81_226}. Measurements of $\mathcal{R}(K^{(*)})$ differed from the SM prediction by about $3.1\sigma$~\cite{NP18_277} previously but this has recently decreased to $0.2\sigma$~\cite{2212.09153}. Since violation of the LFU would be a clear sign of new physics, further tests of the LFU in other SL decays of heavy quarks are well-motivated. The Cabibbo-favored decays $\Lambda_c^+\rightarrow \Lambda \ell^+\nu_{\ell}$, the dominant $\Lambda_c^+$ SL decays~\cite{pdg2020}, provide an ideal platform to test the LFU via the $e$- and $\mu$-modes, by measuring the ratios of their differential decay rates in separate four-momentum transfer regions.
In addition, the angular distribution and the polarization effects in $\Lambda_c^+\rightarrow \Lambda \ell^+\nu_{\ell}$ decays have received more and more attention in recent years~\cite{PRD93_034008,EPJC76_628,PLB792_214,PRD104_013005,JHEP09_208,PRD103_054018,2210.15588,PRD101_094017,PRL118_082001}. 
Various asymmetrical parameters that characterize the angular dependence of the decay distributions are also accessible. 
In particular, the forward-backward asymmetry in the lepton system ($A^{\ell}_{\rm FB}$), which is the least sensitive to uncertainties of the form factors (FFs) parametrizing the hadronic matrix elements, can provide a critical test of LFU~\cite{PRD93_034008}. The forward-backward asymmetry in the hadronic $p\pi^-$ system ($A^{p}_{\rm FB}$) can be used to extract the longitudinal polarization of the $\Lambda$ baryon and the decay asymmetry parameter $\alpha_{\Lambda_c}$~\cite{PLB275_495,PRL94_191801}. Therefore, a model-independent measurement of these angular observables and decay rates is of great interest to test various theoretical calculations~\cite{PRD93_034008,EPJC76_628,PLB792_214,PRD104_013005,JHEP09_208,PRD103_054018,2210.15588,PRD101_094017,PRL118_082001} and LFU.
Furthermore, in the SM, the time-reversal (T) asymmetry in $\Lambda_c^+\rightarrow \Lambda \ell^+\nu_{\ell}$ is zero due to the absence of a weak phase in the $\Lambda_c^+\rightarrow \Lambda$ transition~\cite{2210.15588,PRD106_053006}. 
Hence, measuring the T asymmetry in $\Lambda_c^+\rightarrow \Lambda \ell^+\nu_{\ell}$ is a clean way to search for new physics with a $CP$-violating phase beyond the SM.  Throughout this paper, charge-conjugate modes are implied unless explicitly noted.

This Letter presents an improved measurement of the absolute branching fraction (BF) of $\Lambda_c^+\rightarrow \Lambda \mu^+\nu_{\mu}$~\cite{PLB767_42}. We also report the first measurement of the differential decay rates and the forward-backward asymmetries for $\Lambda_c^+\rightarrow \Lambda \ell^+\nu_{\ell}$ decays in the full kinematic range as well as individual four-momentum transfer regions. Using these observables, we provide the first test on LFU using $\Lambda_c^+\rightarrow \Lambda \ell^+\nu_{\ell}$ decays.
The forward-backward asymmetries are obtained from both the current muonic sample and a re-examination of the previous electronic sample of Ref.~\cite{Lamev}.  
Furthermore, an improved measurement of FFs in $\Lambda_c^+\rightarrow \Lambda \ell^+\nu_{\ell}$ decays in comparison to Ref.~\cite{Lamev} is provided, by analyzing the data of $\Lambda_c^+\rightarrow \Lambda \mu^+\nu_{\mu}$ and $\Lambda_c^+\rightarrow \Lambda e^+\nu_e$ decays simultaneously. The data sets are collected at the center-of-mass energies $\sqrt{s}=4.600$, 4.612, 4.628, 4.641, 4.661, 4.682, and 4.699 GeV, with a total integrated luminosity of $4.5~\mathrm{fb}^{-1}$~\cite{lum_4600,lum_new}. At these energies, no additional hadrons accompanying the $\Lambda^+\bar{\Lambda}_c^-$ pairs are kinematically allowed. 

Details about the BESIII detector design and performance are provided in Refs.~\cite{Ablikim:2009aa,Muon,Ablikim:2019hff}.
Monte Carlo (MC) simulations for the signal and background samples are similar as those described in Ref.~\cite{Lamev}. 
We select “single tag” (ST) and “double tag” (DT) samples to measure the absolute BF of $\Lambda_c^+\rightarrow \Lambda \mu^+\nu_{\mu}$ decay.
Single tag decays are $\bar{\Lambda}^-_c$ baryons reconstructed from their final state particles in one of the fourteen hadronic decays used in Ref.~\cite{pKev}. Double tag decays are events with a ST and a $\Lambda^+_c$ baryon candidate reconstructed as $\Lambda \mu^+\nu_{\mu}$.  
The ST $\bar{\Lambda}^-_c$ signal candidates are identified using the beam
constrained mass
$$M_{\rm BC}=\sqrt{(\sqrt{s}/2)^2-|\vec{p}_{\bar{\Lambda}^-_c}|^2},$$
where $\vec{p}_{\bar{\Lambda}^-_c}$ is the measured momentum of the ST $\bar{\Lambda}^-_c$ candidate. 
A kinematic variable $\Delta E=E_{\rm beam}-E_{\bar{\Lambda}^-_c}$ is required to improve the signal significance for ST $\bar{\Lambda}^-_c$ baryons. 
The $\Delta E$ requirements, $M_{\rm BC}$ distributions and their ST yields are documented in Ref.~\cite{pKev}. 
The total ST yield in all data sets is $N^{\rm ST}=122,268\pm474$, where the uncertainty is statistical only.

Candidates for the signal mode $\Lambda^+_c\rightarrow \Lambda \mu^+\nu_\mu$ are
selected from the remaining tracks recoiling against the ST
$\bar{\Lambda}^-_c$ candidates. The $\Lambda \to p \pi^-$ candidate is selected with the same
criteria as those used in the ST procedures~\cite{Lamev}. 
The $\mu$ candidate is identified with the combined information of $dE/dx$ measured in the multilayer drift chamber, time of flight, and the energy measured in the electromagnetic calorimeter (EMC), and is required to satisfy 
$\mathcal{L}_{\mu} > 0.001$, $\mathcal{L}_{\mu} > \mathcal{L}_{e}$ and
$\mathcal{L}_{\mu}>\mathcal{L}_{K}$, where $\mathcal{L}_{\mu}$, $\mathcal{L}_{e}$ and $\mathcal{L}_{K}$ are the particle identification probabilities for a muon, electron and kaon, respectively. Studies of inclusive MC samples show that the backgrounds are dominated by $\Lambda_c^+\rightarrow
\Lambda\pi^+$, $\Sigma^0\pi^+$ and $\Lambda\pi^+\pi^0$.
Backgrounds from $\Lambda_c^+\rightarrow\Lambda\pi^+$ and $\Lambda_c^+\rightarrow \Sigma^0\pi^+$ are rejected by a requirement on the $\Lambda\mu^+$ invariant mass, $M_{\Lambda\mu^+} < 2.12$ GeV. The background from
$\Lambda_c^+\rightarrow \Lambda\pi^+\pi^0$ is suppressed by requiring the maximum energy of all unused photon clusters, $E_{\gamma \rm{max}}$, to be less than 0.25 GeV and the deposited energy for the muon candidate in the EMC to be less than 0.30 GeV.
Since the neutrino is not detected, we employ the kinematic variable
$U_{\rm miss}=E_{\rm miss}-|\vec{p}_{\rm miss}|$ to obtain information on the neutrino, where $E_{\rm miss}$ and $\vec{p}_{\rm miss}$ are the missing energy and momentum, respectively, carried by the neutrino, as inferred from energy-momentum conservation.  

\begin{figure}[tp!]
\begin{center}
   \begin{minipage}[t]{8cm}
   \includegraphics[width=\linewidth]{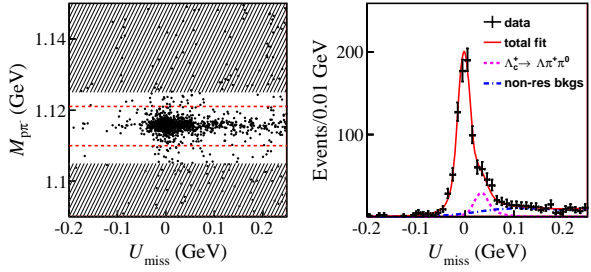}
   \end{minipage}   
   \caption{ (left) The $M_{p\pi^-}$ versus $U_{\rm miss}$ distribution for the $\Lambda_c^+\rightarrow \Lambda \mu^+\nu_{\mu}$
   candidates. The area between the dashed lines denotes the $\Lambda$ signal region and the hatched areas indicate the $\Lambda$ sideband regions. (right) Projected $U_{\rm miss}$ distribution within the $\Lambda$ signal region together with the fit. }
\label{fig:umiss_data_sig}
\end{center}
\end{figure}

Figure~\ref{fig:umiss_data_sig} (left) shows the distribution of
$M_{p\pi^-}$ versus $U_{\rm miss}$ for the $\Lambda^+_c\to \Lambda\mu^+\nu_{\mu}$ candidates in data. A cluster of the events is located around
the intersection of the $\Lambda$ and $\Lambda \mu^+\nu_{\mu}$ signal
region. Requiring $M_{p\pi^-}$ to be within the $\Lambda$ signal
region, we project the distribution onto the $U_{\rm miss}$ axis, as
shown in Fig.~\ref{fig:umiss_data_sig} (right). 
To obtain the number of signal events, the $U_{\rm miss}$ distribution is fitted with a signal function which consists of 
a Gaussian to describe the core of the $U_{\rm miss}$
distribution and two power law tails to account for initial and final state radiations~\cite{PRD79_052010,PRL115_221805}.
A MC-simulated background shape is used to describe the peaking background 
from $\Lambda_c^+\rightarrow \Lambda\pi^+\pi^0$, and a MC-simulated non-resonant background shape is used to describe the continuous background. 
In the fit, the size of peaking background from $\Lambda_c^+\rightarrow \Lambda\pi^+\pi^0$ is fixed according to the BF $\mathcal{B}(\Lambda_c^+\rightarrow \Lambda\pi^+\pi^0)$~\cite{pdg2020} and the simulated efficiency~\cite{JHEP12_033}.  
From the fit, we obtain the yield of $\Lambda_c^+\rightarrow \Lambda \mu^+\nu_{\mu}$ decay $N^{\rm DT}=752\pm31$, where the uncertainty is statistical only.

The absolute BF for $\Lambda_c^+\rightarrow \Lambda \mu^+\nu_{\mu}$ is
determined by
\begin{equation} \mathcal{B}(\Lambda_c^+\rightarrow \Lambda \mu^+\nu_{\mu})=\frac{N^{\rm DT}}{N^{\rm ST}\times\varepsilon_{\rm semi}},
\label{eq:branch}
\end{equation}
where $\varepsilon_{\rm semi}=0.1767$, is the average efficiency for detecting the $\Lambda_c^+\rightarrow \Lambda
\mu^+\nu_{\mu}$ decay in ST events~\cite{PRL115_221805,Lamev}. Inserting the values of $N^{\rm DT}$, $N^{\rm ST}$, $\varepsilon_{\rm semi}$ into Eq.~(\ref{eq:branch}), we measure $\mathcal B({\Lambda^+_c\rightarrow
\Lambda \mu^+\nu_{\mu}})=(3.48\pm0.14\pm0.10)\%$, where the first uncertainty is statistical, and the second is systematic uncertainty described below.

With the DT technique, the BF measurement is insensitive to the systematic uncertainties of the ST selection. The remaining systematic uncertainties in the BF measurement are described as follows. 
The uncertainties of the $\mu^+$ tracking and particle identification efficiencies are determined to be 0.3\% and 0.5\% studied using $e^+e^-\rightarrow \gamma\mu^+\mu^-$ events.
The uncertainty due to $\Lambda$ reconstruction is determined to be 0.2\%, studied with $J/\psi\rightarrow pK^-\bar{\Lambda}$ and $J/\psi\rightarrow \Lambda\bar{\Lambda}$ control samples.
The uncertainty associated with the simulation of the SL signal model is estimated to be 1.0\% by varying the input form-factor parameters by one standard deviations as determined in this work. Further systematic uncertainties include: fit to the
$U_{\rm miss}$ distribution (1.8\%) estimated by using alternative
signal shapes and background shapes, requirements on $M_{\Lambda\mu^+}$ (0.8\%), $E_{\gamma \rm{max}}$ (0.6\%), and deposited energy (0.8\%), 
the quoted BF for $\Lambda\rightarrow p\pi^-$
(0.8\%)~\cite{pdg2020}, the MC sample size (0.8\%), and the ST yield (1.0\%).
Adding these contributions in quadrature gives a total systematic uncertainty of 2.9\% for the $\mathcal{B}(\Lambda_c^+\rightarrow \Lambda \mu^+\nu_{\mu})$ measurement.

\begin{figure}[htbp]
   \begin{center}
   \includegraphics[width=\linewidth]{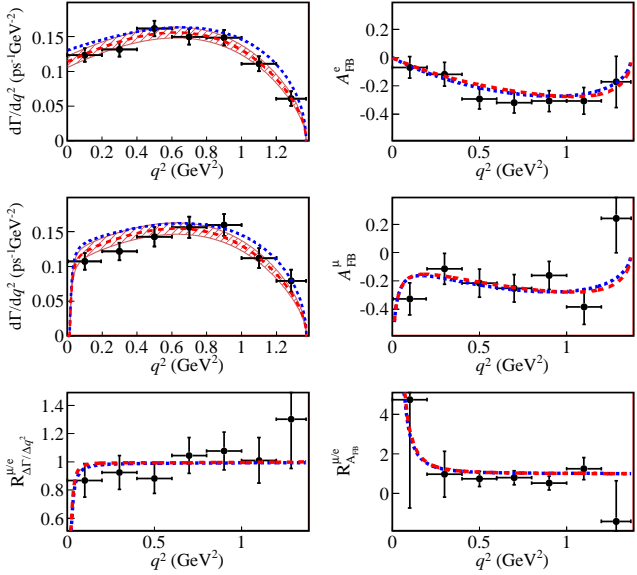}
   \caption{Top Row: Measurements of (left) $\Delta\Gamma/\Delta q^2$ and (right) $A^{\ell}_{\rm FB}$ vs.~$q^2$ for $\Lambda_c^+\rightarrow \Lambda e^+\nu_e$.  
Middle Row: the same quantities for  $\Lambda_c^+\rightarrow \Lambda \mu^+\nu_{\mu}$.   Bottom Row: the ratios of the above, $\mathcal{R}^{\mu/e} $. The dots with error bars are data, where statistical and systematic uncertainties are both included. The red dash-dotted curves show the derived values using the FFs measured in this work, while the blue dashed curves show those predicted by LQCD~\cite{PRL118_082001}. The bands show the total uncertainties of $d\Gamma/d q^2$ using the FFs measured in this work. } 
   \label{fig:Decayrates}
\end{center}
\end{figure}

The partial decay rate of $\Lambda^+_c\rightarrow \Lambda \ell^+\nu_{\ell}$ is measured by analyzing the $\ell^+\nu_{\ell}$ mass-squared ($q^2$) distribution.  The candidates of $\Lambda_c^+\rightarrow \Lambda e^+\nu_e$ are selected as in Ref.~\cite{Lamev}. 
The $q^2$ distributions of $\Lambda^+_c\rightarrow \Lambda \ell^+\nu_{\ell}$ candidates are divided into seven bins of [0,~0.20), [0.20,~0.40), [0.40,~0.60), [0.60,~0.80), [0.80,~1.00), [1.00,~1.20) and [1.20,~$q^2_{\rm max}$)~GeV$^2$, where $q^2_{\rm max}=(M_{\Lambda_c}-M_{\Lambda})^2$ and $M_{\Lambda_{(c)}}$ is the mass of $\Lambda_{(c)}$. 
The measured partial decay rate in the $i^{th}$ $q^2$ bin, $\Delta \Gamma_i$, is determined by 
\begin{equation}
\Delta \Gamma_i=\int_i \frac{d\Gamma}{dq^2} dq^2=\sum\limits_{j=1}^{N_{\rm bins}}(\epsilon^{-1})_{ij}N^j_{\rm DT}/(\tau_{\Lambda_c}\times N^{\rm ST}),
\label{eq:drate}
\end{equation}
where $\tau_{\Lambda_c}$ is the lifetime of $\Lambda^+_c$~\cite{pdg2020} and $\epsilon_{ij}$ is the efficiency matrix to describe the reconstruction efficiency of SL decays and migration effects across $q^2$ bins~\cite{PRL124_241802,PRD101_112002}. 
The SL signal yield observed in the $j^{th}$ $q^2$ bin $N^j_{\rm DT}$ is obtained from a fit to the corresponding $U_{\rm miss}$ distribution. 
The measured differential decay rates, $\Delta\Gamma_i/\Delta q^2$, for decays $\Lambda_c^+\rightarrow \Lambda e^+\nu_e$ and $\Lambda_c^+\rightarrow \Lambda \mu^+\nu_{\mu}$ as well as their ratios $\mathcal{R}^{\mu/e}_{\Delta \Gamma/\Delta q^2}$ in each bin are shown in Fig.~\ref{fig:Decayrates} (left). 
The distributions of the $d\Gamma/dq^2$ and $\mathcal{R}^{\mu/e}_{\Delta \Gamma/\Delta q^2}$ as a function of $q^2$ calculated using the FFs measured in this work (described later) and those predicted by LQCD~\cite{PRL118_082001} are also shown in Fig.~\ref{fig:Decayrates} (left). 
The ratios $\mathcal{R}^{\mu/e}_{\Delta \Gamma/\Delta q^2}$ measured in different $q^2$ intervals are consistent with LQCD calculation within one standard deviation, which suggests that no evidence for violation of LFU is found.

The model-independent forward-backward asymmetries of both the lepton system, $A^{\ell}_{\rm FB}(q^2)$, and of the $p\pi^-$ system, $A^{p}_{\rm FB}(q^2)$ are also measured with 
$\Lambda_c^+\rightarrow \Lambda \ell^+\nu_{\ell}$ candidates. The definition of forward-backward asymmetry $A^{\ell,p}_{\rm FB}(q^2)$ follows~\cite{PRD93_034008,EPJC76_628}
\begin{equation}
A^{\ell,p}_{\rm FB}(q^2)=\frac{\int_{0}^{1}\frac{d^2\Gamma}{dq^2d{\rm cos}\theta_{\ell,p}}d{\rm cos}\theta_{\ell,p}-\int_{-1}^{0}\frac{d^2\Gamma}{dq^2d{\rm cos}\theta_{\ell,p}}d{\rm cos}\theta_{\ell,p}}{\int^{1}_{0}\frac{d^2\Gamma}{dq^2d{\rm cos}\theta_{\ell,p}}d{\rm cos}\theta_{\ell,p}+\int_{-1}^{0}\frac{d^2\Gamma}{dq^2d{\rm cos}\theta_{\ell,p}}d{\rm cos}\theta_{\ell,p}},
\label{eq:AFB}
\end{equation}
where $d^2\Gamma/(dq^2d{\rm cos}\theta_{\ell,p})$ is the two-dimensional differential rate, and $\theta_{\ell}$ is the angle between the lepton ($e^+/\mu^+$) direction and the $\ell^+\nu_{\ell}$ system direction in the 
$\Lambda^+_c$ rest frame, and $\theta_{p}$ is the angle between the proton and the $\Lambda$ direction also in the $\Lambda^+_c$ rest frame.
To measure $A^{\ell,p}_{\rm FB}(q^2)$, the $q^2$ distributions of the SL candidates are separated into two regions of ${\rm cos}\theta_{\ell,p}\in(0,~1)$ and ${\rm cos}\theta_{\ell,p}\in(-1,~0)$.  
With the similar procedure as applied in measuring the partial decay rate defined in Eq.~(\ref{eq:drate}), the $\Delta\Gamma_i/\Delta q^2$ within ${\rm cos}\theta_{\ell,p}\in(0,~1)$ and ${\rm cos}\theta_{\ell,p}\in(-1,~0)$ is measured for each $q^2$ bin. Then we extract the forward-backward asymmetries, $A^{\ell}_{\rm FB}(q^2)$ and $A^{p}_{\rm FB}(q^2)$, for each $q^2$ bin with Eq.~(\ref{eq:AFB}). The results for $A^{\ell}_{\rm FB}(q^2)$ of $\Lambda_c^+\rightarrow \Lambda e^+\nu_e$ and $\Lambda_c^+\rightarrow \Lambda \mu^+\nu_{\mu}$ as well as their ratios $\mathcal{R}^{\mu/e}_{A_{\rm FB}}$ are shown in Fig.~\ref{fig:Decayrates} (right), where their dependences calculated with the FFs measured in this work and those predicted by LQCD~\cite{PRL118_082001} are also presented. Measurements on $\mathcal{R}^{\mu/e}_{A_{\rm FB}}$ in various $q^2$ bins also show no evidence for a violation of the LFU.  
\normalsize

\normalsize
\begin{figure}[htbp]
   \begin{center}
   \begin{minipage}[t]{8cm}   
   \includegraphics[width=\linewidth]{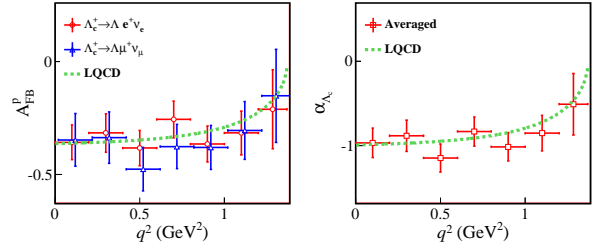}
   \end{minipage}
   \caption{Measurements of (left) $A^{p}_{\rm FB}(q^2)$ for $\Lambda_c^+\rightarrow \Lambda \ell^+\nu_{\ell}$ and (right) the averaged $\alpha_{\Lambda_c}$ of the $e$- and $\mu$-modes in each  $q^2$ interval. The dots with error bars are data, where statistical and systematic uncertainties are both included. The dashed curve show the derived values predicted by LQCD~\cite{PRL118_082001}. In drawing the blue dots, the bin-center value of $q^2$ is shifted manually by 0.02~GeV$^{2}$ to avoid overlap. } 
   \label{fig:alphaLc}
\end{center}
\end{figure}

The $A^{p}_{\rm FB}(q^2)$ for $\Lambda_c^+\rightarrow \Lambda e^+\nu_e$ and $\Lambda_c^+\rightarrow \Lambda \mu^+\nu_{\mu}$ measured in each $q^2$ interval are shown in Fig.~\ref{fig:alphaLc} (left).  Using the relation
\begin{equation}
\alpha_{\Lambda_c}(q^2)=\frac{2}{\alpha_{\Lambda}} \left[ A^{p}_{\rm FB}(q^2) \right],
\end{equation}
where $\alpha_{\Lambda}$ is the $\Lambda\rightarrow p\pi^-$ decay asymmetry parameter~\cite{pdg2020}, the model-independent decay asymmetry parameter $\alpha_{\Lambda_c}$ for $\Lambda_c^+\rightarrow \Lambda \ell^+\nu_{\ell}$ in separate $q^2$ intervals are measured. Differences in  
$\alpha_{\Lambda_c}$ between the $e$- and $\mu$-modes are expected at a order of $10^{-3}$~\cite{PRD93_034008,EPJC76_628,2210.15588}, indistinguishable with current statistics. Alternatively, their averaged value is determined in each $q^2$ interval as shown  in Fig.~\ref{fig:alphaLc} (right). This is the first model-independent determination of $\alpha_{\Lambda_c}$ as a function of $q^2$ distribution. In addition, we determined the value averaged over $q^2$ to be $\left \langle \alpha_{\Lambda_c}\right \rangle=-0.94\pm0.07_{\rm stat.}\pm0.03_{\rm syst.}$.

We also measure the T asymmetry parameter $\mathcal{T}_p$ defined by~\cite{2210.15588}:
\begin{equation}
\mathcal{T}_p=\frac{\left[\left(\int_{-\pi}^{0}-\int_0^{\pi}\right)d\chi\right]\left[\left(\int_{0}^{1}-\int_{-1}^{0}\right)d{\rm cos}\theta_p\right]\Gamma^{\ell}_{\chi,{\rm cos}\theta_p}}{\alpha_{\Lambda}\Gamma^{\ell}}, \nonumber
\end{equation}
where $\Gamma^{\ell}$ is the total decay rate and $\Gamma^{\ell}_{\chi,{\rm cos}\theta_p}$ is the two-dimensional decay rate as a function of $\chi$ and ${\rm cos}\theta_p$ distributions in $\Lambda_c^+\rightarrow \Lambda \ell^+\nu_{\ell}$, where $\chi$ is the acoplanarity angle between the $\Lambda$ and $W^+$ decay planes. We measure $\mathcal{T}_p(\Lambda_c^+\rightarrow \Lambda e^+\nu_e)=-0.021\pm0.041_{\rm stat.}\pm0.001_{\rm syst.}$ and $\mathcal{T}_p(\Lambda_c^+\rightarrow \Lambda \mu^+\nu_{\mu})=0.068\pm0.055_{\rm stat.}\pm0.002_{\rm syst.}$. The two results are consistent with zero, as predicted from the SM~\cite{2210.15588}, with no indication of new physics in $\Lambda_c^+\rightarrow \Lambda \ell^+\nu_{\ell}$ decays.

To improve experimental precision of the FF parameters in the $\Lambda^+_c\rightarrow \Lambda$ transition, the candidates from $\Lambda_c^+\rightarrow \Lambda \mu^+\nu_{\mu}$ in this work and $\Lambda^+_c\rightarrow \Lambda e^+\nu_{e}$ obtained in Ref.~\cite{Lamev} are analyzed simultaneously.
The differential decay rate of $\Lambda^+_c\rightarrow \Lambda \ell^+\nu_{\ell}$ depends on four kinematic variables: $q^2$, helicity angles $\theta_p$ and $\theta^{\prime}_{\ell}$~\cite{angle}, and the acoplanarity angle $\chi$. The differential decay rate described in terms of helicity amplitudes $H_{\lambda_{\Lambda}\lambda_{W}}$ is~\cite{PLB275_495,PRL94_191801,EPJC59_27}
\small
\begin{eqnarray}
&& \frac{d^4\Gamma}{dq^2d{\rm cos}\theta^{\prime}_{\ell}d{\rm cos}\theta_{p}d\chi} = \frac{G^2_F|V_{cs}|^2}{2(2\pi)^4}\cdot\frac{Pq^2(1-m^2_{\ell}/q^2)^2}{24M^2_{\Lambda_c}} \times ~~~~~~~~~~~~~~~ \nonumber \\
&&~~\left\{\frac{3}{8}(1-{\rm cos}\theta^{\prime}_{\ell})^2|H_{\frac{1}{2}1}|^2(1+\alpha_{\Lambda} {\rm cos}\theta_{p}) \right. \nonumber \\
&&~+\frac{3}{8}(1+{\rm cos}\theta^{\prime}_{\ell})^2|H_{-\frac{1}{2}-1}|^2(1-\alpha_{\Lambda} {\rm cos}\theta_{p})  \nonumber \\
&&~+\frac{3}{4}{\rm sin}^2\theta^{\prime}_{\ell}\left[|H_{\frac{1}{2}0}|^2(1+\alpha_{\Lambda}{\rm cos}\theta_{p})+|H_{-\frac{1}{2}0}|^2(1-\alpha_{\Lambda}{\rm cos}\theta_{p})\right] \nonumber \\
&&~+\frac{3}{2\sqrt{2}}\alpha_{\Lambda}{\rm cos}\chi {\rm sin}\theta^{\prime}_{\ell} {\rm sin}\theta_{p}\times \nonumber \\
&&~~~~~\left.\left[(1-{\rm cos}\theta^{\prime}_{\ell})H_{-\frac{1}{2}0}H_{\frac{1}{2}1}+(1+{\rm cos}\theta^{\prime}_{\ell})H_{\frac{1}{2}0}H_{-\frac{1}{2}-1}\right]+\mathcal{H}_{m^2_{\ell}}\right\},
\label{eq:decayrate}
\end{eqnarray}
\normalsize
where $m_{\ell}$ is the lepton mass. The definitions of the variables $H_{\lambda_{\Lambda}\lambda_{W}}$, $G_F$, $|V_{cs}|$, and $P$ follows those in Ref.~\cite{Lamev}.
The detailed expression of $\mathcal{H}_{m^2_{\ell}}$ is documented in the supplementary materials~\cite{supple}. 

The helicity amplitudes of vector and axial-vector components, $H^{V}_{\lambda_{\Lambda}\lambda_{W}}$ and $H^{A}_{\lambda_{\Lambda}\lambda_{W}}$, parameterized in Eq.~(\ref{eq:decayrate}) are related to six FFs by~\cite{PRL118_082001}
\begin{eqnarray}
H^{V/A}_{\frac{1}{2}1}&=&\sqrt{2Q_{\mp}}f_{\bot}/g_{\bot}(q^2),  \nonumber \\
H^{V/A}_{\frac{1}{2}0}&=&\sqrt{Q_{\mp}/q^2}f_{+}/g_{+}(q^2)(M_{{\it \Lambda}_c}\pm M_{{\it \Lambda}}), \nonumber \\
H^{V/A}_{\frac{1}{2}t}&=&\sqrt{Q_{\pm}/q^2}f_{0}/g_{0}(q^2)(M_{{\it \Lambda}_c}\mp M_{{\it \Lambda}}),
\end{eqnarray}
where the scalar helicity component is denoted by $\lambda_W=t$ and $Q_{\pm}=(M_{\Lambda_c}\pm M_{\Lambda})^2-q^2$. 
The contributions of FFs $f_0(q^2)$ and $g_0(q^2)$ are very small in decay rate and difficult to be determined.
Hence the shapes of $f_0(q^2)$ and $g_0(q^2)$ are fixed according to a LQCD calculation~\cite{PRL118_082001}. 
The definitions of the independent free parameters formed in the other four form factors, $f_{\bot,+}(q^2)$ and $g_{\bot,+}(q^2)$, follow Ref.~\cite{Lamev}. 
An additional constraint, $g_+(q^2_{\max})=g_{\perp}(q^2_{\rm max})$, introduced by endpoint relations for baryons~\cite{JHEP11_073} is  taken into account, which
results in $g_{+}(q^2)= g_{\perp}(q^2)$ as we neglect the differences of the slope parameters ($\alpha_1^{g_{+,\bot}}$) between $g_{+}(q^2)$ and $g_{\perp}(q^2)$. 
Therefore, the five independent free parameters, $a_0^{g_{\perp}}$, $\alpha^{g_{\perp}}_1$, $\alpha^{f_{\perp}}_1$, $r_{f_{+}}=a_0^{f_{+}}/a_0^{g_{\perp}}$, $r_{f_{\perp}}=a_0^{f_{\perp}}/a_0^{g_{\perp}}$, are introduced to describe the FFs.

\begin{figure}[htbp]
\begin{center}
   \begin{minipage}[t]{8cm}
   \includegraphics[width=\linewidth]{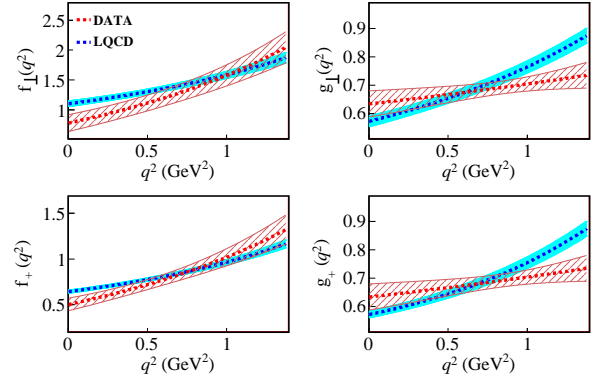}
   \end{minipage}
   \caption{Comparison of form factors with LQCD calculations. The bands show the total uncertainties.} 
   \label{fig:form_Lev}
\end{center}
\end{figure}

A four-dimensional maximum likelihood fit is performed to the variables $q^2$, $\cos\theta^{\prime}_{\ell}$, $\cos\theta_p$, and $\chi$ within the $U_{\rm miss}$ signal regions defined by $(-0.06,0.06)$~GeV for both $\Lambda^+_c\rightarrow \Lambda e^+\nu_{e}$ and $\Lambda^+_c\rightarrow \Lambda \mu^+\nu_{\mu}$ decays. In order to determine the parameter $a_0^{g_{\perp}}$ from the fit, the differential decay rate of $\Lambda_c^+\rightarrow \Lambda \mu(e)^+\nu_{\mu(e)}$ is constrained by the BF of $\Lambda_c^+\rightarrow \Lambda \mu(e)^+\nu_{\mu(e)}$ measured in this work (Ref.~\cite{Lamev}) and the 
$\Lambda^+_c$ lifetime~\cite{pdg2020} using the relation
\begin{equation}
\int\limits^{q^2_{\rm max}}_{m^2_{\ell}} \frac{d\Gamma(\Lambda_c^+\rightarrow \Lambda \ell^+\nu_{\ell})}{dq^2}dq^2=\frac{\mathcal{B}(\Lambda_c^+\rightarrow \Lambda \ell^+\nu_{\ell})}{\tau_{\Lambda_c}}.
\label{eq:a0fperp}
\end{equation}
The projections of the fit onto $q^2$, $\cos\theta^{\prime}_{\ell}$, $\cos\theta_p$, and $\chi$, as well as the fitted form-factor parameters of $a_0^{g_{\perp}}$, $\alpha_1^{g_{\perp}}$, $\alpha^{f_{\perp}}_1$, $r_{f_+}$ and $r_{f_{\perp}}$ are shown in the supplementary material~\cite{supple}. 
Using a treatment similar to Ref.~\cite{Lamev}, the systematic uncertainties in $a_0^{g_{\perp}}$, $\alpha_1^{g_{\perp}}$, $\alpha^{f_{\perp}}_1$, $r_{f_+}$, and $r_{f_{\perp}}$ are estimated to be 0.8\%, 6.3\%, 2.2\%, 1.3\%, and 5.8\%, respectively.  In comparison to Ref.~\cite{Lamev}, the precision in measuring $a_0^{g_{\perp}}$, $\alpha_1^{g_{\perp}}$, $r_{f_+}$ and $r_{f_{\perp}}$ is improved by 25\%, 11\%, 22\% and 43\%, respectively. The comparisons to the LQCD calculations~\cite{PRL118_082001} are shown in Fig.~\ref{fig:form_Lev}. The dependence of the measured FFs show different kinematic behavior compared to the LQCD calculations, especially for the slopes of these four FFs. 

In summary, we report an improved measurement of the absolute BF of the
$\Lambda^+_c\rightarrow \Lambda \mu^+\nu_{\mu}$ decay, 
$\mathcal B({\Lambda^+_c\rightarrow \Lambda
\mu^+\nu_{\mu}})=(3.48\pm0.14\pm0.10)\%$, by analyzing 4.5~fb$^{-1}$ data
collected at center-of-mass energies 
ranging from 4.600~GeV to 4.699~GeV at BESIII. 
This work supersedes our previous measurement~\cite{PLB767_42} and improves the precision of
the world average value by a factor of three. Comparisons to the theoretical predictions are also shown in Table \ref{tab:theorybf}. 
The predicted the BFs in Refs.~\cite{PRD93_034008,PRC72_035201} differ by more than two standard deviations from the measured
$\mathcal B({\Lambda^+_c\rightarrow \Lambda \mu^+\nu_{\mu}})$. Thus, our measurement disfavors these predictions at a confidence level of more than $95\%$. 
Combining with $\mathcal{B}(\Lambda_c^+\rightarrow \Lambda e^+\nu_e)=(3.56\pm0.11\pm0.07)\%$ in Ref.~\cite{Lamev}, we determine the ratio
$\mathcal{B}(\Lambda^+_c\rightarrow \Lambda
\mu^+\nu_{\mu})/\mathcal{B}(\Lambda^+_c\rightarrow \Lambda
e^+\nu_{e})=0.98\pm0.05_{\rm stat.}\pm0.03_{\rm syst.}$, which is consistent with the value 0.97 as predicted by the LQCD. 
The ratio of the two BFs in different four-momentum transfer
regions is also studied, and no evidence for the LFU violation is found. 
Combining the measured BFs of $\Lambda^+_c\rightarrow \Lambda \mu^+\nu_{\mu}$ and $\Lambda^+_c\rightarrow \Lambda e^+\nu_{e}$~\cite{Lamev}, $\tau_{\Lambda_c}$~\cite{pdg2020} and the $q^2$-integrated rate predicted by LQCD~\cite{PRL118_082001}, we determine $|V_{cs}|=0.937\pm0.014_{\mathcal{B}}\pm0.024_{\rm LQCD}\pm0.007_{\tau_{\Lambda_c}}$, which is 
the most precise determination obtained with the charmed baryon SL decays. This result is consistent with $|V_{cs}|=0.972\pm0.007$ measured in $D\rightarrow K\ell\nu_{\ell}$ decays~\cite{pdg2020} within 1.2 standard deviations.

In addition, the model-independent lepton forward-backward asymmetries $A^{e,\mu}_{\rm FB}(q^2)$ as well as their ratios are measured for the first time. Their average values are determined to be $\left \langle A^{e}_{\rm FB}\right \rangle=-0.24\pm0.03_{\rm stat.}\pm0.01_{\rm syst.}$ and $\left \langle A^{\mu}_{\rm FB}\right \rangle=-0.22\pm0.04_{\rm stat.}\pm0.01_{\rm syst.}$, respectively. The measured $\left \langle A^e_{\rm FB}\right \rangle$ and $\left \langle A^{\mu}_{\rm FB}\right \rangle$ are consistent with the predictions in Refs.~\cite{PRD93_034008,EPJC76_628,2210.15588,PRL118_082001}, but clearly differ from the predictions in Ref.~\cite{PRD104_013005}. The detailed comparisons are shown in Tab.~\ref{tab:theorybf}.
Investigations on $\mathcal{R}^{\mu/e}_{A_{\rm FB}}$ in different four-momentum transfer regions show no evidence for the LFU violation. As a first measurement of the forward-backward asymmetries in the $p\pi^-$ system, we determine the averaged values $\left \langle A^{p}_{\rm FB}\right \rangle=-0.33\pm0.03_{\rm stat.}\pm0.01_{\rm syst.}$ and $\left \langle A^{p}_{\rm FB}\right \rangle=-0.37\pm0.04_{\rm stat.}\pm0.01_{\rm syst.}$ for $\Lambda^+_c\rightarrow \Lambda e^+\nu_{e}$ and $\Lambda^+_c\rightarrow \Lambda \mu^+\nu_{\mu}$, respectively. Ignoring the negligible predicted difference between the $e$- and $\mu$-modes, the $\left \langle \alpha_{\Lambda_c}\right \rangle$ is determined to be $-0.94\pm0.07_{\rm stat.}\pm0.03_{\rm syst.}$, which is consistent with the theoretical predictions in Refs.~\cite{PRD93_034008,EPJC76_628,PLB792_214,PRD104_013005,JHEP09_208,PRD103_054018,2210.15588,PRD101_094017,PRL118_082001} and the model-dependent measurement by CLEO~\cite{PRL94_191801}. 
Our measurements of the ${\rm T}$ asymmetry parameters, $\mathcal{T}_p(\Lambda_c^+\rightarrow \Lambda e^+\nu_e)=-0.021\pm0.041_{\rm stat.}\pm0.001_{\rm syst.}$ and $\mathcal{T}_p(\Lambda_c^+\rightarrow \Lambda \mu^+\nu_{\mu})=0.068\pm0.055_{\rm stat.}\pm0.002_{\rm syst.}$,  are consistent with zero as predicted in the SM.
Furthermore, our results on the form factors provide more stringent tests of and calibration for LQCD calculations of $\Lambda_c^+\rightarrow \Lambda \ell^+\nu_{\ell}$ decays, which are important for both charmed baryon decays~\cite{2103.07064,PLB823_13675} and $\Lambda_b$ decays~\cite{PRD104_013005,PRD80_096007,PRD90_114033,PRD85_014035,PRD88_014512,PRD92_034503,PRD93_074501,2107.13140}. 

\begin{table}
\caption{ Comparisons of $\mathcal{B}(\Lambda_c^+\rightarrow \Lambda \mu^+\nu_{\mu})$ [in \%], $\left \langle \alpha_{\Lambda_c}\right \rangle$, $\left \langle A^{e}_{\rm FB}\right \rangle$ and $\left \langle A^{\mu}_{\rm FB}\right \rangle$ from theories and measurement. } 
\begin{center}
\resizebox{!}{2.8cm}{
\begin{tabular}
{l|cccc} \hline\hline 
 &  $\mathcal{B}(\Lambda_c^+\rightarrow \Lambda \mu^+\nu_{\mu})$ & $\left \langle \alpha_{\Lambda_c}\right \rangle$ & $\left \langle A^{e}_{\rm FB}\right \rangle$ & $\left \langle A^{\mu}_{\rm FB}\right \rangle$  \\ \hline
CQM~\cite{PRD93_034008}     & 2.69             & $-0.87$ & $-0.2$     & $-0.21$ \\
RQM~\cite{EPJC76_628}            & 3.14           & $-0.86$ & $-0.209$ & $-0.242$ \\
CQM(HONR)~\cite{PRC72_035201} & 4.25     &  &  & \\ 
NRQM~\cite{PRD95_053005}     & 3.72 & & & \\
HBM~\cite{2210.15588} & $3.67\pm0.23$ & $-0.826$ & $-0.176(5)$ & $-0.143(6)$ \\
LQCD~\cite{PRL118_082001}     & $3.69\pm0.22$ & $-0.874(10)$ & $-0.201(6)$ & $-0.169(7)$ \\
LCSR~\cite{PRD80_074011}       & $3.0\pm0.3$ & &  \\
$SU(3)$~\cite{PLB792_214}                   & $3.6\pm0.4$ & $-0.86(4)$ & &  \\      
LFCQM~\cite{PRD101_094017} & $3.21\pm0.85$ &  $-0.97(3)$ & & \\
MBM~\cite{PRD101_094017} & 3.38  & $-0.83$ & &  \\
LFQM~\cite{PRD104_013005} & $3.90\pm0.73$ & $-0.87(9)$ & $0.20(5)$ & $0.16(4)$ \\
LFCQM~\cite{PRD103_054018} & $3.40\pm1.02$ & $-0.97(3)$ & & \\
$SU(3)$~\cite{PLB823_136765} & $3.45\pm0.30$ & & & \\
This work                                     & $3.48\pm0.17$ & $-0.94(8)$ & $-0.24(3)$ & $-0.22(4)$ \\
\hline\hline
\end{tabular}
}
\label{tab:theorybf}
\end{center}
\end{table}

The BESIII Collaboration thanks the staff of BEPCII and the IHEP computing center for their strong support. This work is supported in part by National Key R\&D Program of China under Contracts Nos. 2020YFA0406400, 2020YFA0406300; National Natural Science Foundation of China (NSFC) under Contracts Nos. 11635010, 11735014, 11835012, 11935015, 11935016, 11935018, 11961141012, 12022510, 12025502, 12035009, 12035013, 12061131003, 12192260, 12192261, 12192262, 12192263, 12192264, 12192265; the Chinese Academy of Sciences (CAS) Large-Scale Scientific Facility Program; the CAS Center for Excellence in Particle Physics (CCEPP); Joint Large-Scale Scientific Facility Funds of the NSFC and CAS under Contract No. U1832207; CAS Key Research Program of Frontier Sciences under Contracts Nos. QYZDJ-SSW-SLH003, QYZDJ-SSW-SLH040; 100 Talents Program of CAS; The Institute of Nuclear and Particle Physics (INPAC) and Shanghai Key Laboratory for Particle Physics and Cosmology; ERC under Contract No. 758462; European Union's Horizon 2020 research and innovation programme under Marie Sklodowska-Curie grant agreement under Contract No. 894790; German Research Foundation DFG under Contracts Nos. 443159800, 455635585, Collaborative Research Center CRC 1044, FOR5327, GRK 2149; Istituto Nazionale di Fisica Nucleare, Italy; Ministry of Development of Turkey under Contract No. DPT2006K-120470; National Research Foundation of Korea under Contract No. NRF-2022R1A2C1092335; National Science and Technology fund of Mongolia; National Science Research and Innovation Fund (NSRF) via the Program Management Unit for Human Resources \& Institutional Development, Research and Innovation of Thailand under Contract No. B16F640076; Polish National Science Centre under Contract No. 2019/35/O/ST2/02907; The Royal Society, UK under Contract No. DH160214; The Swedish Research Council; U. S. Department of Energy under Contract No. DE-FG02-05ER41374.

\clearpage
\appendix
\onecolumngrid
\section*{\boldmath Supplementary Materials of Study of $\Lambda_c^+\rightarrow \Lambda \mu^+\nu_{\mu}$ and Test of Lepton Flavor Universality with $\Lambda_c^+\rightarrow \Lambda \ell^+\nu_{\ell}$ Decays }

The term $\mathcal{H}_{m^2_{\ell}}$ shown in the differential decay rate of $\Lambda^+_c\rightarrow \Lambda \ell^+\nu_{\ell}$ is described in terms of helicity amplitudes $H_{\lambda_{\Lambda}\lambda_{W}}$, including the vector ($H^{V}_{\lambda_{\Lambda}\lambda_{W}}$) and axial-vector ($H^{V}_{\lambda_{\Lambda}\lambda_{W}}$) components. The $\mathcal{H}_{m^2_{\ell}}$ depends on the following kinematic variables: 
the $\ell^+\nu_{\ell}$ mass-squared ($q^2$), the angle between the proton momentum in the $\Lambda$ rest frame and the $\Lambda$ momentum in the $\Lambda^+_c$ rest frame ($\theta_p$),
the angle between the lepton ($e^+/\mu^+$) momentum in the $W^+$ rest frame and the $\Lambda$ momentum in the $\Lambda^+_c$ rest frame ($\theta^{\prime}_{\ell}$), and the acoplanarity angle between the $\Lambda$ and $W^+$ decay planes ($\chi$). It is given by~\cite{PLB275_495,PRL94_191801,EPJC59_27}:
\begin{eqnarray}
\mathcal{H}_{m^2_{\ell}}&=&\frac{m^2_{\ell}}{2q^2} \left[ \frac{3}{2}|H_{\frac{1}{2}t}|^2(1+\alpha_{{\Lambda}} {\rm cos}\theta_{p})+\frac{3}{2}|H_{-\frac{1}{2}t}|^2(1-\alpha_{{\Lambda}} {\rm cos}\theta_{p}) 
+3{\rm cos}\theta^{\prime}_{\ell} \left[H_{\frac{1}{2}t}H_{\frac{1}{2}0}(1+\alpha_{{\Lambda}} {\rm cos}\theta_{p})+H_{-\frac{1}{2}t}H_{-\frac{1}{2}0}(1-\alpha_{{\Lambda}} {\rm cos}\theta_{p}) \right] \right. \nonumber \\
&+&\frac{3}{2}{\rm cos}^2\theta^{\prime}_{\ell} \left[|H_{\frac{1}{2}0}|^2(1+\alpha_{{\Lambda}} {\rm cos}\theta_{p})+|H_{-\frac{1}{2}0}|^2(1-\alpha_{{\Lambda}} {\rm cos}\theta_{p})\right]
+\frac{3}{4}{\rm sin}^2\theta^{\prime}_{\ell} \left[|H_{\frac{1}{2}1}|^2(1+\alpha_{{\Lambda}} {\rm cos}\theta_{p})+|H_{-\frac{1}{2}-1}|^2(1-\alpha_{{\Lambda}} {\rm cos}\theta_{p})\right] \nonumber \\ 
&+& \left. \frac{3}{\sqrt{2}}\alpha_{{\Lambda}}{\rm cos}\chi {\rm sin}\theta^{\prime}_{\ell}{\rm sin}\theta_{p} (H_{-\frac{1}{2}t}H_{\frac{1}{2}1}-H_{\frac{1}{2}t}H_{-\frac{1}{2}-1}) 
+\frac{3}{\sqrt{2}}\alpha_{{\Lambda}}{\rm cos}\chi {\rm sin}\theta^{\prime}_{\ell}{\rm cos}\theta^{\prime}_{\ell} {\rm sin}\theta_{p} (H_{-\frac{1}{2}0}H_{\frac{1}{2}1}-H_{\frac{1}{2}0}H_{-\frac{1}{2}-1}) \right],
\label{eq:decayrate}
\end{eqnarray}
\normalsize
where the scalar helicity component is denoted by $\lambda_W=t$, $H_{\lambda_{\Lambda}\lambda_{W}}=H^V_{\lambda_{\Lambda}\lambda_{W}}-H^A_{\lambda_{\Lambda}\lambda_{W}}$ and $H^{V(A)}_{-\lambda_{\Lambda}-\lambda_{W}}=+(-)H^{V(A)}_{\lambda_{\Lambda}\lambda_{W}}$, and $\alpha_{\Lambda}$ is the $\Lambda\rightarrow p\pi^-$ decay asymmetry parameter~\cite{pdg2020}.

\begin{figure}[htbp]
\begin{center}
   \includegraphics[width=\linewidth]{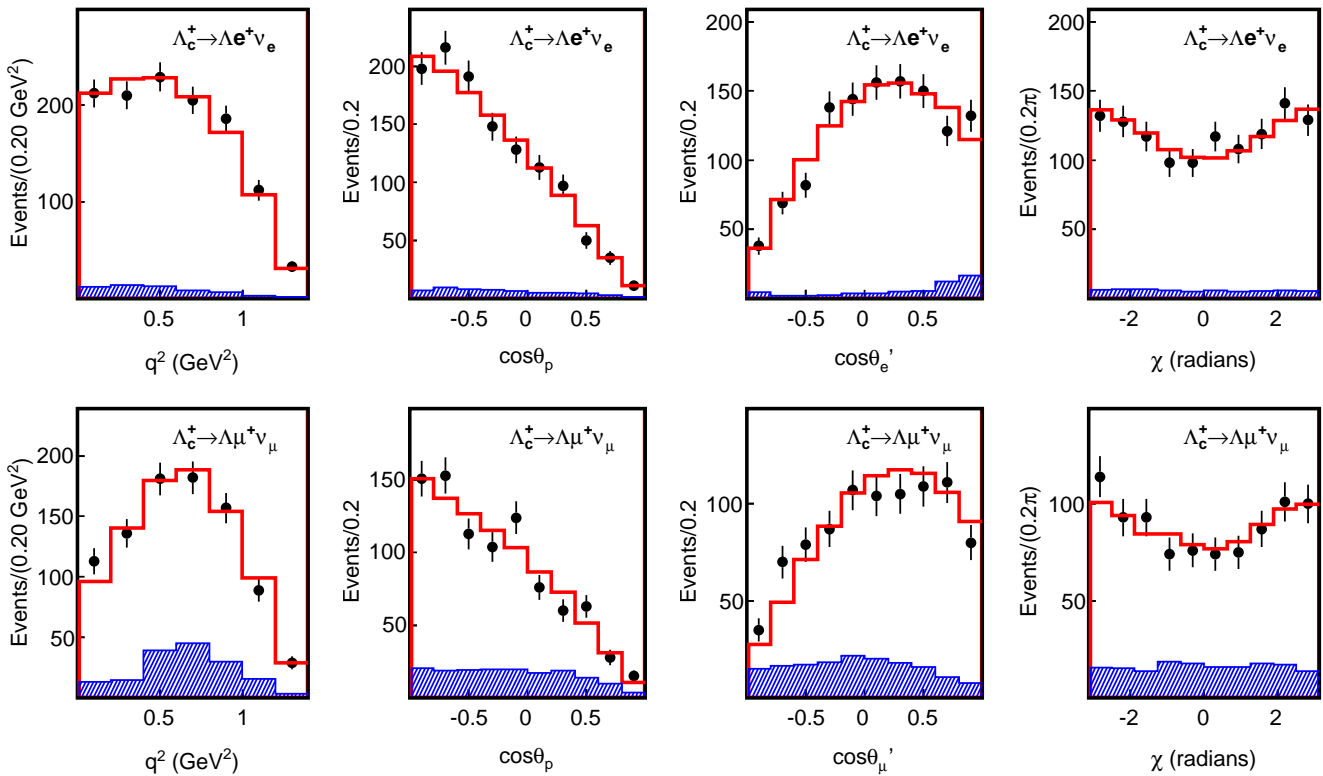}
   \caption{ Projections of the fitted kinematic variables for comparison in  data (dots with error bars) and MC. The solid histograms are the MC-simulated total distributions including signal plus background components. The shadowed histograms show the distributions of the simulated background. 
The top (bottom) row shows $\Lambda_c^+\rightarrow \Lambda e^+\nu_e$ 
($\Lambda_c^+\rightarrow \Lambda \mu^+\nu_{\mu}$). }
   \label{fig:Form}
\end{center}
\end{figure}

Figure~\ref{fig:Form} shows the fit projections onto the $q^2$, $\cos\theta_p$, $\cos\theta^{\prime}_{e/\mu}$ and $\chi$ distributions for decays $\Lambda_c^+\rightarrow \Lambda e^+\nu_e$ and $\Lambda_c^+\rightarrow \Lambda \mu^+\nu_\mu$, respectively. The fitted form-factor parameters of $a_0^{g_{\perp}}$, $\alpha_1^{g_{\perp}}$, $\alpha^{f_{\perp}}_1$, $r_{f_+}$, and $r_{f_{\perp}}$ are given in Tab.~\ref{tab:finalresults}. 

\begin{table*}
\begin{center}
\caption{Measured form-factor parameters, where the first errors are statistical and the second are systematic. The lower table shows the correlation coefficients of statistical and systematic uncertainties between the form factor parameters. } \normalsize
\begin{tabular}
{|c|c|c|c|c|c|c|c|c|} \hline   Parameters & $a_0^{g_{\perp}}$  & $\alpha^{g_{\perp}}_1$ & $\alpha^{f_{\perp}}_1$ & $r_{f_+}$ & $r_{f_{\perp}}$ \\
\hline Values &  $0.57\pm0.03\pm0.01$ & $1.64\pm1.86\pm0.10$ & $-6.42\pm1.62\pm0.14$ & $1.61\pm0.25\pm0.02$ & $2.49\pm0.37\pm0.14$  \\
\hline 
\hline Coefficients  & & $\alpha^{g_{\perp}}_1$ & $\alpha^{f_{\perp}}_1$ & $r_{f_+}$ & $r_{f_{\perp}}$  \\
\hline 
 $a_0^{g_{\perp}}$         &                &    $-0.88$ & ~~~$0.69$   & $-0.79$     &  $-0.75$ \\
 $\alpha^{g_{\perp}}_1$ &                &                 & $-0.71$  &   ~~~$0.61$     &  ~~~$ 0.71$ \\
 $\alpha^{f_{\perp}}_1$  &                &                 &               & $-0.85$     &  $-0.70$ \\
 $r_{f_{+}}$                    &                &                &                 &                 & ~~~$0.66$ \\
\hline
\end{tabular}
\label{tab:finalresults}
\end{center}
\end{table*}

%


\begin{thebibliography}{99}
\bibitem{CKM} M. Kobayashi and T. Maskawa, \href{https://doi.org/10.1016/j.nuclphysbps.2005.01.049}{Prog. Theor. Phys. {\bf 49}, 652 (1973)}.
\bibitem{pdg2020} R.~L.~Workman $et~al.$ (Particle Data Group), \href{https://pdg.lbl.gov/}{Prog. Theor. Exp. Phys. {\bf 2022}, 083C01 (2022)}.
\bibitem{PRL109_101802} J.~P.~Lees $et~al.$ [BABAR Collaboration], \href{https://doi.org/10.1103/PhysRevLett.109.101802}{Phys. Rev. Lett. {\bf 109}, 101802 (2012)}.
\bibitem{PRD88_072012} J.~P.~Lees $et~al.$ [BABAR Collaboration], \href{https://doi.org/10.1103/PhysRevD.88.072012}{Phys. Rev. D {\bf 88}, 072012 (2013)}.
\bibitem{PRD92_072014} M.~Huschle $et~al.$ [Belle Collaboration], \href{https://doi.org/10.1103/PhysRevD.92.072014}{Phys. Rev. D {\bf 92}, 072014 (2015)}.
\bibitem{PRD94_072007} Y.~Sato $et~al.$ [Belle Collaboration], \href{https://doi.org/10.1103/PhysRevD.94.072007}{Phys. Rev. D {\bf 94}, 072007 (2016)}. 
\bibitem{PRL115_159901} R.~Aaij $et~al.$ [LHCb Collaboration], \href{https://doi.org/10.1103/PhysRevLett.115.111803}{Phys. Rev. Lett. {\bf 115}, 111803 (2015)}.
\bibitem{PRL118_21101} S.~Hirose $et~al.$ [Belle Collaboration], \href{https://doi.org/10.1103/PhysRevLett.118.211801}{Phys. Rev. Lett. {\bf 118}, 211801 (2017)}. 
\bibitem{PRD97_072013} R.~Aaij $et~al.$ [LHCb Collaboration], \href{https://doi.org/10.1103/PhysRevD.97.072013}{Phys. Rev. D {\bf 97}, 072013 (2018)}.
\bibitem{PRL120_171802} S.~Hirose $et~al.$ [Belle Collaboration], \href{https://doi.org/10.1103/PhysRevLett.120.171802}{Phys. Rev. Lett. {\bf 120}, 171802 (2018)}. 

\bibitem{PRD86_032012} J.~P.~Lees $et~al.$ [BABAR Collaboration], \href{https://doi.org/10.1103/PhysRevD.86.032012}{Phys. Rev. D {\bf 86}, 032012 (2012)}.
\bibitem{JHEP06_108} R.~Aaij $et~al.$ [LHCb Collaboration], \href{https://doi.org/10.1007/JHEP06(2017)108}{JHEP {\bf 06}, 108 (2017)}.
\bibitem{JHEP08_055} R.~Aaij $et~al.$ [LHCb Collaboration], \href{https://doi.org/10.1007/JHEP08(2017)055}{JHEP {\bf 08}, 055 (2017)}.
\bibitem{PRL122_191801} R.~Aaij $et~al.$ [LHCb Collaboration], \href{https://doi.org/10.1103/PhysRevLett.122.191801}{Phys. Rev. Lett. {\bf 122}, 191801 (2019)}.
\bibitem{JHEP05_040} R.~Aaij $et~al.$ [LHCb Collaboration], \href{https://doi.org/10.1007/JHEP05(2020)040}{JHEP {\bf 05} 040, (2020)}.
\bibitem{JHEP03_105} S.~Choudhury $et~al.$ [Belle Collaboration], \href{https://doi.org/10.1007/JHEP03(2021)105}{JHEP {\bf 03} 105, (2021)}
\bibitem{NP18_277} R.~Aaij $et~al.$ [LHCb Collaboration], \href{https://doi.org/10.1038/s41567-021-01478-8}{Nat. Phys. {\bf 18}, 277 (2022)}.
\bibitem{EPJC81_226} Y.~Amhis $et~al.$ [HFLAV Collaboration], \href{https://doi.org/10.1140/epjc/s10052-020-8156-7}{Eur. Phys. J. C {\bf 81}, 226 (2021)}.
\bibitem{2212.09153} R.~Aaij $et~al.$ [LHCb Collaboration], \href{https://doi.org/10.48550/arXiv.2212.09152}{arXiv:2212.09152}; \href{https://doi.org/10.48550/arXiv.2212.09153}{arXiv:2212.09153}. 

\bibitem{PRD93_034008} T.~Gutsche, M.~A.~Ivanov, J.G. K{\"o}rner, V.~E.~Lyubovitskij and P.~Santorelli, \href{https://doi.org/10.1103/PhysRevD.93.034008}{Phys. Rev. D {\bf 93}, 034008 (2016)}.
\bibitem{EPJC76_628} R.~N. Faustov, V.~O.~Galkin, \href{https://doi.org/10.1140/epjc/s10052-016-4492-z}{Eur. Phys. J. C {\bf 76} 628 (2016)}.
\bibitem{PRD104_013005} Y.~S.~Li, X.~Liu and F.~S.~Yu, \href{https://doi.org/10.1103/PhysRevD.104.013005}{Phys. Rev. D {\bf 104}, 013005 (2021)}.
\bibitem{JHEP09_208} M.~Golz, G.~Hiller and T.~Magorsch, \href{https://doi.org/10.1007/JHEP09(2021)208}{JHEP {\bf 09}, 208 (2021)}.
\bibitem{2210.15588} C.~Q.~Geng, X.~N.~Jin and C.~W.~Liu, \href{https://doi.org/10.1103/PhysRevD.107.033008}{Phys. Rev. D {\bf 107}, 033008 (2023)}.

\bibitem{PLB792_214} C.~Q.~Geng, C.~W.~Liu, T.~H.~Tsai, S.~W.~Yeh, \href{https://doi.org/10.1016/j.physletb.2019.03.056}{Phys. Lett. B {\bf 792}, 214 (2019)}.
\bibitem{PRD103_054018} C.~Q.~Geng, C.~W.~Liu and T.~H.~Tsai, \href{https://doi.org/10.1103/PhysRevD.103.054018}{Phys. Rev. D {\bf 103}, 054018 (2021)}.
\bibitem{PRD101_094017} C.~Q.~Geng, Chong-Chung Lih, C.~W.~Liu, and T.~H.~Tsai, \href{https://doi.org/10.1103/PhysRevD.101.094017}{Phys. Rev. D {\bf 101}, 094017 (2020)}.
\bibitem{PRL118_082001} S.~Meinel, \href{https://doi.org/10.1103/PhysRevLett.118.082001}{Phys. Rev. Lett. {\bf 118}, 082001 (2017)}. 
\bibitem{PLB275_495} J.G. K{\"o}rner and M. Kr{\"a}mer, \href{https://doi.org/10.1016/0370-2693(92)91623-H}{Phys. Lett. B {\bf 275}, 495 (1992)}.
\bibitem{PRL94_191801} J.~W.~Hinson $et$ $al$. [CLEO Collaboration], \href{https://doi.org/10.1103/PhysRevLett.94.191801}{Phys. Rev. Lett. {\bf 94}, 191801 (2005)}.


\bibitem{PRD106_053006} C.~Q.~Geng, X.~N.~Jin and C.~W.~Liu, \href{https://doi.org/10.1103/PhysRevD.106.053006}{Phys. Rev. D {\bf 106}, 053006 (2022)}.
\bibitem{PLB767_42} M.~Ablikim $et~al.$ [BESIII Collaboration], \href{https://doi.org/10.1016/j.physletb.2017.01.047}{Phys. Lett. B {\bf 767}, 42 (2017)}.

\bibitem{Lamev} M.~Ablikim $et~al.$ [BESIII Collaboration], \href{https://doi.org/10.1103/PhysRevLett.129.231803}{Phys. Rev. Lett. {\bf 129}, 231803 (2022)}. 


\bibitem{lum_4600} M.~Ablikim $et~al.$ [BESIII Collaboration], \href{https://doi.org/10.1088/1674-1137/39/9/093001}{Chin. Phys. C {\bf 39}, 093001 (2015)}.
\bibitem{lum_new} M.~Ablikim $et~al.$ [BESIII Collaboration], \href{https://doi.org/10.1088/1674-1137/ac84cc}{Chin. Phys. C {\bf 46}, 113003 (2022)}.

\bibitem{Ablikim:2009aa} M.~Ablikim $et~al.$ [BESIII Collaboration], \href{https://doi.org/10.1016/j.nima.2009.12.050}{Nucl.\ Instrum.\ Meth.\ A {\bf 614}, 345 (2010)}.
\bibitem{Muon} K.~X.~Huang $et~al.$, \href{https://doi.org/10.1007/s41365-022-01133-8}{Nucl. Sci. Tech. {\bf 33}, 142 (2022).} 
\bibitem{Ablikim:2019hff} M.~Ablikim $et~al.$ [BESIII Collaboration], \href{https://doi.org/10.1088/1674-1137/44/4/040001}{Chin. Phys. C {\bf 44}, 040001 (2020)}.

\bibitem{pKev} M.~Ablikim $et~al.$ [BESIII Collaboration],  \href{https://doi.org/10.1103/PhysRevD.106.112010}{Phys. Rev. D {\bf 106}, 112010 (2022)}.

\bibitem{PRL115_221805} M.~Ablikim $et~al.$ [BESIII Collaboration], \href{https://doi.org/10.1103/PhysRevLett.115.221805}{Phys. Rev. Lett. {\bf 115}, 221805 (2015)}.
\bibitem{PRD79_052010} J.~Y.~Ge $et~al.$ [CLEO Collaboration], \href{https://doi.org/10.1103/PhysRevD.79.052010}{Phys. Rev. D {\bf 79}, 052010 (2009)}.

\bibitem{JHEP12_033} M.~Ablikim $et~al.$ [BESIII Collaboration], \href{https://doi.org/10.1007/JHEP12(2022)033}{JHEP {\bf 12}, 033 (2022)}.

\bibitem{PRL124_241802} M.~Ablikim $et~al.$ [BESIII Collaboration],  \href{https://doi.org/10.1103/PhysRevLett.124.241802}{Phys. Rev. Lett. {\bf 124}, 241802 (2020)}. 
\bibitem{PRD101_112002} M.~Ablikim $et~al.$ [BESIII Collaboration],  \href{https://doi.org/10.1103/PhysRevD.101.112002}{Phys. Rev. D {\bf 101}, 112002 (2020)}. 

\bibitem{angle} It should be noted that $\theta^{\prime}_{\ell}=\pi-\theta_{\ell}$.
\bibitem{EPJC59_27} A.~Kadeer, J.~G.~K{\"o}rner and U.~Moosbrugger, \href{https://doi.org/10.1140/epjc/s10052-008-0801-5}{Eur. Phys. J. C {\bf 59}, 27 (2009)}.

\bibitem{supple} See Supplemental Material at http://link.aps.org/xxx for the detailed expression for $\mathcal{H}_{m_{\ell}^2}$, projection plots of the kinematic variables for  
$\Lambda_c^+\rightarrow \Lambda e^+\nu_e$ and $\Lambda_c^+\rightarrow \Lambda \mu^+\nu_{\mu}$ decays, and the fitted form-factor parameters from the maximum likelihood fits.

\bibitem{JHEP11_073} G.~Hiller and R.~Zwicky, \href{https://doi.org/10.1007/JHEP11(2021)073}{JHEP {\bf 11}, 073 (2021)}.

\bibitem{PRC72_035201} M. Pervin, W. Roberts and S. Capstick, \href{https://doi.org/10.1103/PhysRevC.72.035201}{Phys. Rev. C {\bf 72}, 035201 (2005)}. 
\bibitem{PRD95_053005} M.~M.~Hussain and W.~Roberts, \href{https://doi.org/10.1103/PhysRevD.95.053005}{Phys. Rev. D {\bf 95}, 053005 (2017)}; \href{https://doi.org/10.1103/PhysRevD.95.099901}{Phys. Rev. D {\bf 95}, 099901 (2017)}.
\bibitem{PRD80_074011} Y.~L.~Liu, M.~Q.~Huang and D.~W.~Wang, \href{https://doi.org/10.1103/PhysRevD.80.074011}{Phys. Rev. D {\bf 80}, 074011 (2009)}.

\bibitem{PLB823_136765} X.~G.~He, F.~Huang, W.~Wang, Z.~P.~Xing, \href{https://doi.org/10.1016/j.physletb.2021.136765}{Phys. Lett. B {\bf 823}, 136765 (2021)}.

\bibitem{2103.07064} Q.~A.~Zhang, H.~Hua, F.~Huang, R.~Li, Y.~Li, C.~D.~L{\"u}, P.~Sun, W.~Wang and Y.~B.~Yang, \href{https://doi.org/10.1088/1674-1137/ac2b12}{Chin. Phys. C {\bf 46}, 011002 (2022)}.
\bibitem{PLB823_13675} X.~G.~He, F.~Huang, W.~Wang, Z.~P.~Xing, \href{https://doi.org/10.1016/j.physletb.2021.136765}{Phys. Lett. B {\bf 823}, 136765 (2021)}.

\bibitem{2107.13140} S.~Meinel and G.~Rendon, \href{https://doi.org/10.1103/PhysRevD.105.054511}{Phys. Rev. D {\bf 105}, 054511 (2022)}.

\bibitem{PRD80_096007} K.~Azizi, M.~Bayar, Y.~Sarac and H.~Sundu, \href{https://doi.org/10.1103/PhysRevD.80.096007}{Phys. Rev. D {\bf 80}, 096007 (2009)}.
\bibitem{PRD90_114033} T.~Gutsche, M.~A.~Ivanov, J.G. K{\"o}rner, V.~E.~Lyubovitskij and P.~Santorelli, \href{https://doi.org/10.1103/PhysRevD.90.114033}{Phys. Rev. D {\bf 90}, 114033 (2014)};   \href{https://doi.org/10.1103/PhysRevD.94.059902}{Phys. Rev. D {\bf 94}, 059902(E) (2016)}.

\bibitem{PRD85_014035} T.~Feldmann and M.~W.~Y.~Yip, \href{https://doi.org/10.1103/PhysRevD.85.014035}{Phys. Rev. D {\bf 85}, 014035 (2012)}; \href{https://doi.org/10.1103/physrevd.86.079901}{Phys. Rev. D {\bf 86}, 079901(E) (2012)}.
\bibitem{PRD88_014512} W.~Detmold, C.~J.~David Lin, S.~Meinel and M.~Wingate, \href{https://doi.org/10.1103/PhysRevD.88.014512}{Phys. Rev. D {\bf 88}, 014512 (2013)}.
\bibitem{PRD92_034503} W.~Detmold, C.~Lehner, and S. Meinel, \href{https://doi.org/10.1103/PhysRevD.92.034503}{Phys. Rev. D {\bf 92}, 034503 (2015)}.
\bibitem{PRD93_074501} W. Detmold and S. Meinel, \href{https://doi.org/10.1103/PhysRevD.93.074501}{Phys. Rev. D {\bf 93}, 074501 (2016)}.

\end{thebibliography}
\end{document}